\documentclass[12pt]{iopart}

\usepackage{graphicx}

\begin{document}

\title{Viscosity and the Soft Ridge at RHIC}
\author{Sean Gavin and George Moschelli}

\address{Physics Department, Wayne State University, 666 W. Hancock, Detroit, MI 48202
}

\begin{abstract}
Correlation studies exhibit a ridge-like feature in rapidity and azimuthal angle, with and without a jet trigger. We ask whether the feature in untriggered correlations can be a consequence of transverse flow and viscous diffusion.  
\end{abstract}

Jet correlation measurements display a ridge of enhanced particle emission that is broad in pseudorapidity $\eta$ and narrow in azimuthal angle $\phi$ \cite{Putschke:2007mi}. Many have attributed this  `hard ridge' to the passage of the jet through the high-density environment produced by the nuclear collision; see e.g. \cite{hwa}. Interestingly, a similar ridge feature is exhibited in untriggered correlation studies where no jet tag is present \cite{Daugherity:2006hz}.  As with the hard ridge, the width of the untagged `soft ridge' broadens in $\eta$ and narrows in $\phi$ for collisions of higher centrality (smaller impact parameter).   

We ask whether the $(\eta, \phi)$ dependence of the soft ridge in untagged correlations can be a consequence of radial flow and viscous diffusion. The connection between radial flow and momentum correlations has been stressed elsewhere \cite{Voloshin:2003ud, Shuryak:2007fu, Pruneau:2007ua}. 
In ref.~\cite{gavin06} we attributed the rapidity  broadening of untriggered momentum correlations to  viscous diffusion, extracting an estimate of the viscosity-to-entropy  ratio in the range from $0.08$ to $0.32$.  In this paper we show that viscous diffusion together with flow can explain the centrality dependence of the soft ridge; see fig.~1 (right). 

The soft ridge has been observed both in momentum fluctuation studies \cite{starMom} and multiplicity correlation measurements \cite{starDense}. Here, we focus on the momentum correlation results, which we argue depend strongly on the viscosity of the high density liquid produced in collisions. Our hope is to eventually reduce the rather large uncertainty in the viscosity estimate in  \cite{gavin06} by using the full landscape of $\eta-\phi$ correlations, rather than just the $\eta$ dependence.    
In this paper, we first recall the role of shear viscosity in the evolution of momentum fluctuations. We survey the hydrodynamic formulation of \cite{gavin06}, extending it to include transverse flow.  We then confront the soft ridge.  

Central nuclear collisions produce a fluid that flows outward with an average transverse 
velocity $v_\perp$. In the hydrodynamic description of these collisions, we typically assume
that $v_\perp$ varies smoothly with spacetime $(t, \mathbf x)$
and is the same for all collisions of a fixed impact parameter. More
realistically, small deviations relative to $v_\perp$  occur throughout the fluid, 
varying with each collision event. Such deviations occur, e.g., 
because the number and location of nucleon-nucleon subcollisions varies in each event.

Viscous friction arises as neighboring elements of the fluid flow past one another. 
This friction drives the stress-energy tensor toward 
the local average $\langle T_{0r} \rangle =  \gamma^2(\epsilon + p)v_\perp$, 
where the energy density is $\epsilon$, the pressure is $p$, and 
$\gamma = (1-v^2)^{-1/2}$.  The extent to which variations  $g_t(\mathbf{x}) 
= T_{0r} - \langle T_{0r}\rangle$ survive the collision depends on the 
magnitude of the viscosity and the lifetime of the fluid.
Near a point where $v_\perp = 0$, the momentum current 
$g_t$ satisfies a diffusion equation, 
$\partial g_t/ \partial t = \nu\nabla^2 g_t $,
where the kinematic viscosity is $\nu = \eta/(\epsilon + p)$ \cite{gavin06}.  To 
incorporate effects of radial and elliptic flow, we assume Bjorken longitudinal 
flow in the $z$ direction coupled to a radial flow ${\mathbf v}_\perp$ in the transverse 
plane that depends only on $\mathbf{r}_\perp$ and $\tau =(t^2 - z^2)^{1/2}$. We write
\begin{equation}
{{\partial g_t} \over{\partial\tau}}
+ {\mathbf v}_\perp\cdot{\mathbf \nabla}_\perp g_t
+ {\mathbf g_t}\cdot {\mathbf \nabla}_\perp v_\perp
= 
\nu
\left({{1}\over{\tau^2}}{{\partial ^{2}}\over{\partial \eta^2}} + \nabla_\perp^2\right) g_t,
\label{eq:temp2}\end{equation} 
where we take $v_\perp \ll 1$ for simplicity.  

Momentum density fluctuations and their dissipation by viscosity are characterized by the 
correlation function
\begin{equation}
r_g = \langle g_t(\mathbf{x}_1)g_t(\mathbf{x}_2)\rangle - \langle
g_t(\mathbf{x}_1)\rangle \langle g_t(\mathbf{x}_2)\rangle,
\label{CcorrF2}
\end{equation}
where the brackets represent an average over an ensemble of events. 
The difference $\Delta r_g$ of $r_g$ from its local equilibrium value $r_{g,\,\rm eq}$ 
also satisfies a diffusion equation \cite{gavin06}.     
We show in ref.~\cite{gavin06} that these momentum density correlations are observable by the relation   
\begin{equation}\label{Cdef}
   {\cal C} =  \langle N\rangle^{-2}\langle \sum_{i\neq j} p_{ti}p_{tj}\rangle -\langle
   p_t\rangle^2 = \langle N\rangle^{-2}\int  \Delta r_g(\mathbf{x_1}, \mathbf{x_2})
   dx_1dx_2,
\end{equation}
where  $i$ labels particles from each event and
$\langle p_t\rangle \equiv \langle \sum p_{ti}\rangle/\langle
N\rangle$. 

The aim is then to reconstruct $\Delta r_g$ by measuring $\cal C$ in different windows of rapidity and azimuthal angle; we emphasized the role of rapidity in determining the viscosity in \cite{gavin06}. While measurements of $\cal C$ would be ideal, much can be learned from  $p_t$ correlation measurements at RHIC and SPS that used somewhat different  observables.  STAR measures the quantity
\begin{equation}
\Delta\sigma_{p_t:n}^2=
\langle N\rangle^{-1}\langle \sum_{i\neq j} (p_{ti}-\langle p_t\rangle)(p_{tj}-\langle p_t\rangle) \rangle;
\end{equation}
CERES and STAR also measure the related quantity $\langle \Delta p_{t1}\Delta p_{t2}\rangle = 
\langle N\rangle  \Delta\sigma_{p_t:n}^2/\langle N(N-1)\rangle$ 
\cite{Adamova:2008sx, Adams:2005ka, Gavin:2003cb}.
Our covariance $\cal C$ is sensitive to the variation of the  $p_t$ of particles as well as their number density, since both quantities effect the momentum current density.  In contrast,  
$\Delta\sigma_{p_t:n}^2$ and $\langle \Delta p_{t1}\Delta p_{t2}\rangle$ are designed to minimize the number density contribution. These quantities are related by  ${\langle N\rangle}^{-1}\Delta\sigma_{p_t:n}^2= 
{\cal C} -\langle p_t\rangle^2{\cal R}$, where  ${\cal R} = (\langle N^2\rangle-\langle N\rangle^2 - 
\langle N\rangle)/\langle N\rangle^2$ measures the number density contribution.

\begin{figure}
\hspace{1pc}%
\begin{minipage}{15pc}
\begin{center}
\includegraphics[width=2.7in]{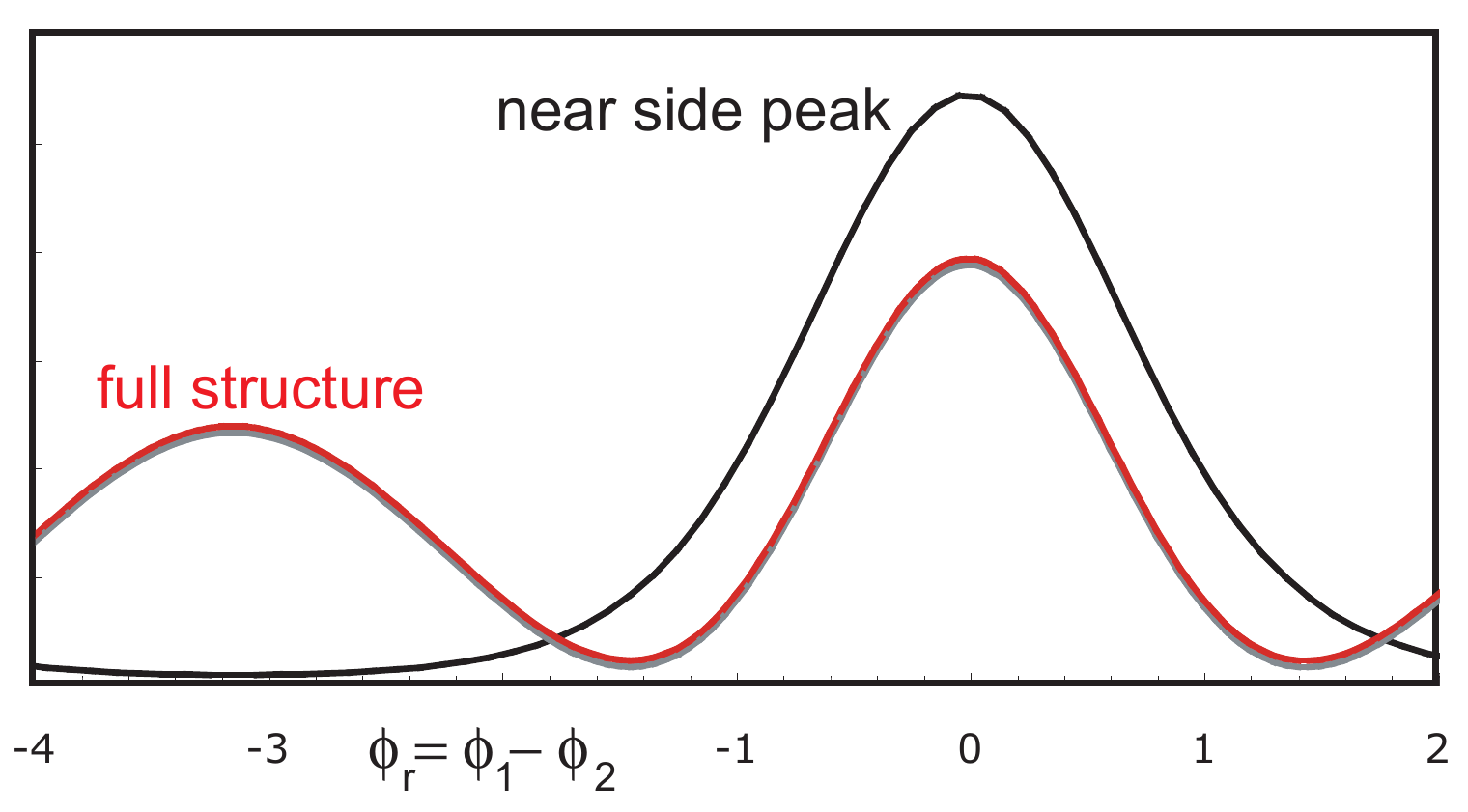} \label{fig:1a}
\end{center}
\end{minipage}\hspace{2pc}%
\begin{minipage}{15pc}
\begin{center}
\includegraphics[width=3.0in]{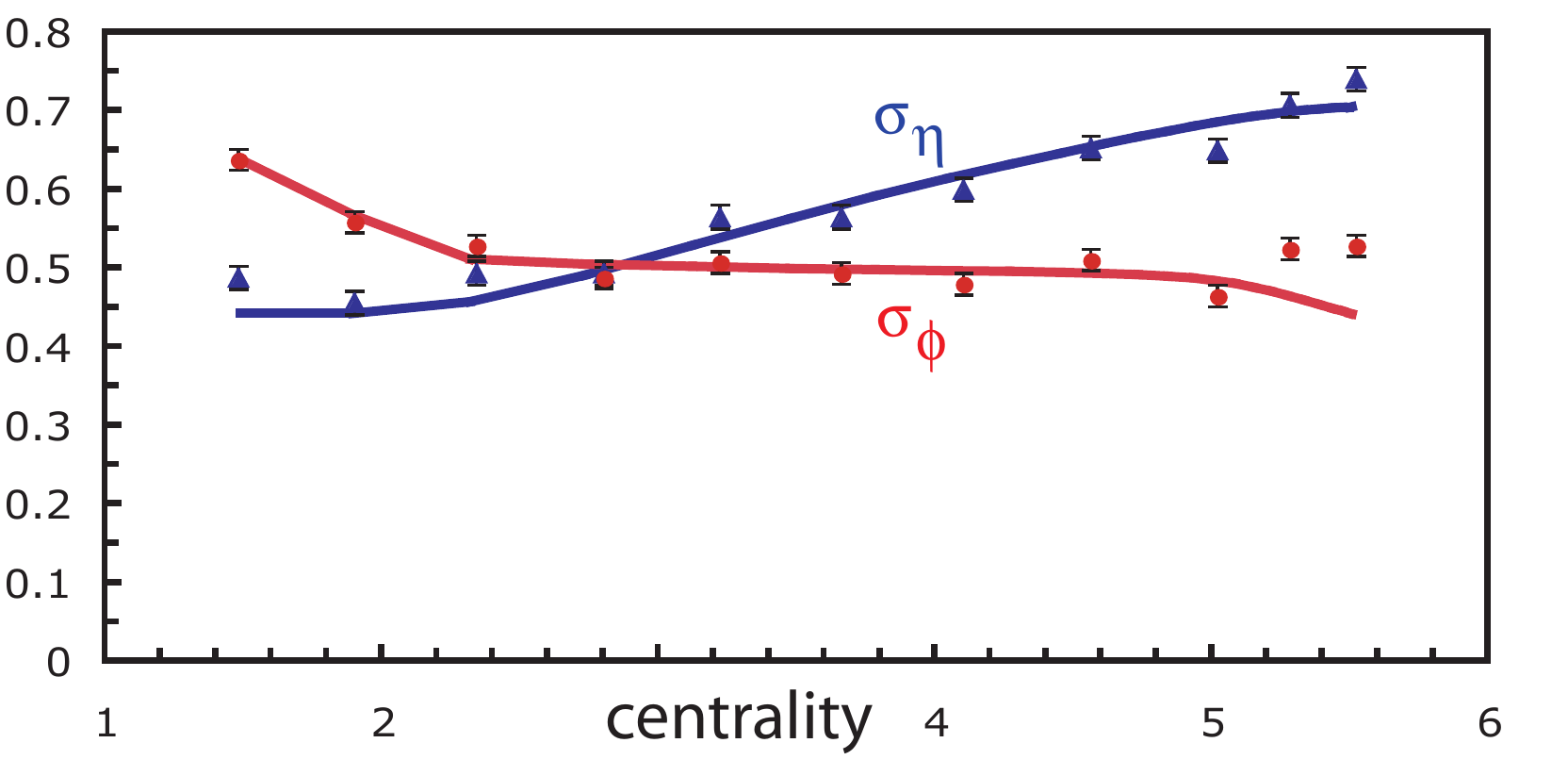}  \label{fig:1b}
\end{center}
\end{minipage}
\vspace{-1.5pc}
\caption{(Left) Schematic correlations in relative azimuthal angle $\phi_r = \phi_1-\phi_2$. (Right) Centrality dependence of the width of the near side peak in $\phi_r$ and $\eta_r = \eta_1-\eta_2$ as a function of centrality. Centrality is determined using $2N_{bin}/N_{part}$ as in \cite{starMom}.}
\end{figure}

STAR employs $\Delta\sigma_{p_t:n}^2$ to construct a correlation function as a
function of  rapidity and azimuthal angle. In essence, they measure $\Delta\sigma_{p_t:n}^2$ in various $\eta_r$ and $\phi_r$ bins, and use the data to construct a correlation function $F(\eta_r, \phi_r)$ defined by  $\Delta\sigma_{p_t:n}^2=\int F d\eta_r d\phi_r$; see \cite{starMom} for details. They find a near-side ridge peaked at $\phi_r = 0$ together with other features attributable to momentum conservation, elliptic flow, and perhaps other phenomena. 

Our primary focus is the near-side peak in azimuth -- the soft ridge -- as shown in fig.~1 (left).  Figure 1 (right) shows the widths of this peak in azimuth $\sigma_\phi$ and rapidity $\sigma_\eta$ as functions of centrality \cite{starMom}. Centrality is measured using the combination $2N_{bin}/N_{part}$, where $N_{bin}$ and $N_{part}$ are, respectively, the numbers of binary collisions and participants. The increase of $\sigma_\eta$ with centrality was used in \cite{gavin06} to estimate the shear viscosity. Diffusion increases the width of $r_g$ relative to its value in peripheral collisions by an amount
%
$\sigma_{\eta,\, c}^2 - \sigma_{\eta,\, p}^2= 4\nu (\tau_{f,\, p}^{-1} -
\tau_{f,\,c}^{-1})$,
%
where $\tau_{f,\,c}$ and $\tau_{f,\,p}$ are the freeze out times in central and peripheral collisions. Transverse flow does not appreciably alter $\sigma_\eta$, since $v_\perp$ is only weakly dependent on $\eta$. Note that the large uncertainty range for the viscosity in \cite{gavin06} follows from the unknown number density contribution relating $\Delta\sigma_{p_t:n}^2$ to $\cal C$.

We turn now to consider the role of viscous diffusion and transverse flow in determining the width in the azimuthal angle as a function of centrality in fig.~1 (right). To see why flow is important, we consider a blast wave model of the mean flow. In a central collision, $\mathbf{v}_\perp = \lambda \mathbf{r}_\perp$. 
A fluid cell a distance $r_\perp$ from the center of the collision volume has a mean speed $v_\perp$ prior to freeze out. Correspondingly, the opening angle into which two particles from this cell emerge is $\phi_r\sim v_{th}/v_\perp \propto (\lambda r_\perp)^{-1}$, for a thermal velocity  $v_{th}\sim 1$.  This introduces correlations, since particles near the center of the collision volume have a large opening angle, while those from a larger $r_\perp$ have a smaller $\phi_r$. A similar correlation may affect the hard ridge as well \cite{Pruneau:2007ua}.

To compute the angular correlations that result from flow and diffusion, we write the momentum correlation function 
\begin{equation}
\Delta r_g(\mathbf{p}_1, \mathbf{p}_2) = 
\int \Delta r_g(\mathbf{x}_1, \mathbf{x}_2)  {{f(\mathbf{x}_1,\mathbf{p}_1)}\over n(\mathbf{x}_1)} 
{{f(\mathbf{x}_2,\mathbf{p}_2)}\over{n(\mathbf{x}_2})}
{{p_1^\mu d\sigma_{1,\,\mu}}\over {(2\pi)^3}} 
 {{p_2^\nu d\sigma_{2,\,\nu}}\over {(2\pi)^3}} 
\label{eq:MomCorr}
\end{equation}
where  $f(\mathbf{x}_a,\mathbf{p}_a)$ is the Boltzmann distribution, $n(\mathbf{x}_a)$ is the density, and the integral is over the Cooper-Frye freeze out surface $\sigma_\mu$. 
Solution of the diffusion equation for $\Delta r_g(\mathbf{x}_1, \mathbf{x}_2)$ derived from (\ref{eq:temp2}) yields the form
\begin{equation}
\Delta r_g(\mathbf{x}_1, \mathbf{x}_2)
\propto \exp\{-r_\perp^2/2\sigma^2 -R_\perp^2/2\Sigma^2\}
\label{eq:SpaceCorr}
\end{equation}
where $\mathbf{r}_\perp = \mathbf{r}_{1\perp}-\mathbf{r}_{2\perp}$ and $\mathbf{R}_\perp = (\mathbf{r}_{1\perp}+\mathbf{r}_{2\perp})/2$. 

We obtain the azimuthal width $\sigma_\phi$ by integrating (\ref{eq:MomCorr}) using (\ref{eq:SpaceCorr}). We use (\ref{eq:temp2}) to compute the spatial widths $\Sigma(\tau)$ and $\sigma(\tau)$ as functions of the proper time $\tau$. The effects of elliptic and radial flow are included 
by assuming an eccentric blast wave form $\mathbf{v}_\perp = \epsilon_x x \hat{x} + \epsilon_y y\hat{y}$. The strength of flow is determined by fitting the measured $\langle p_t\rangle$ and elliptic flow coefficient $v_2$ as in fig.~2. 
\begin{figure}
\hspace{1pc}%
\begin{minipage}{15pc}
\begin{center}
\includegraphics[width=2.9in]{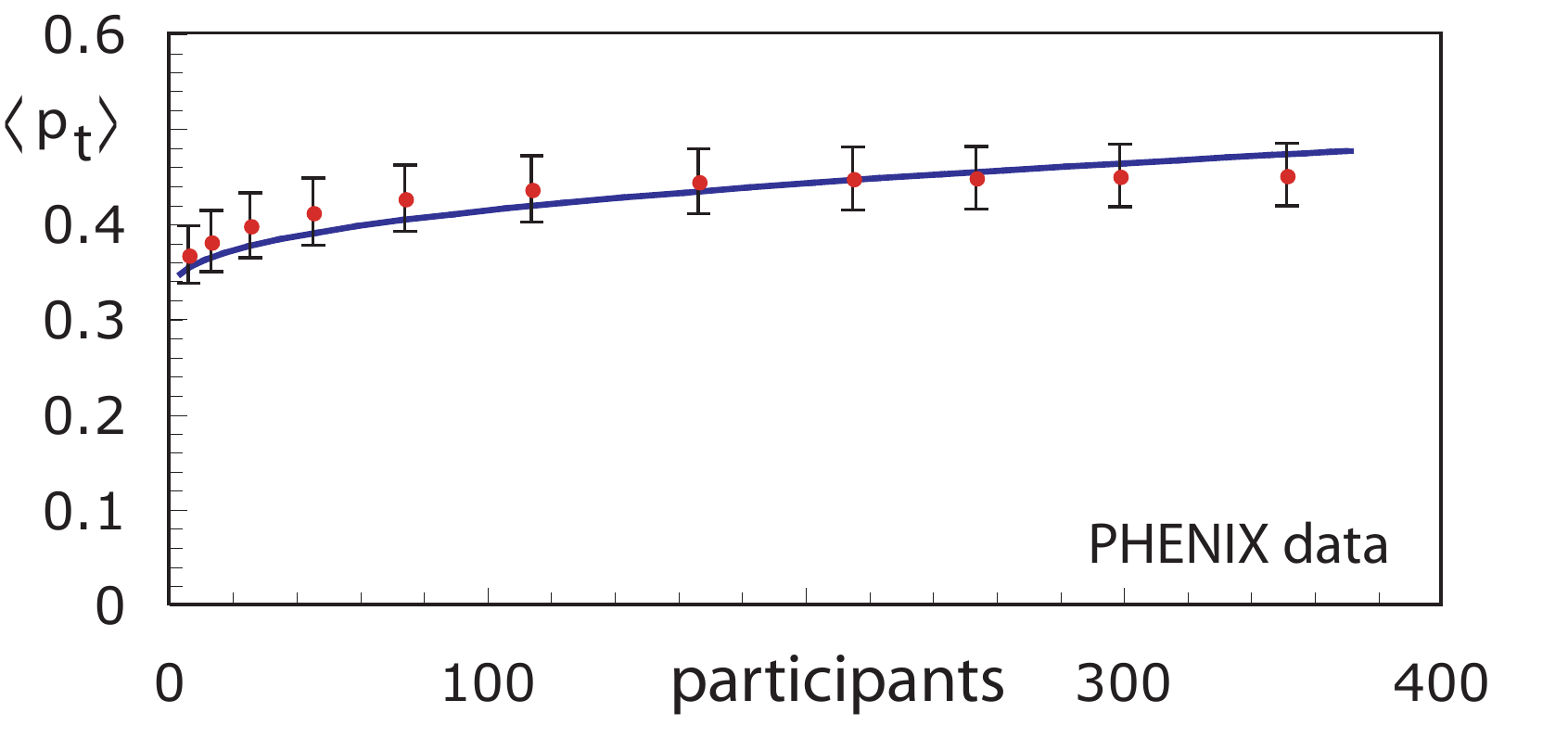} \label{fig:1a}
\end{center}
\end{minipage}\hspace{2pc}%
\begin{minipage}{15pc}
\begin{center}
\includegraphics[width=3.0in]{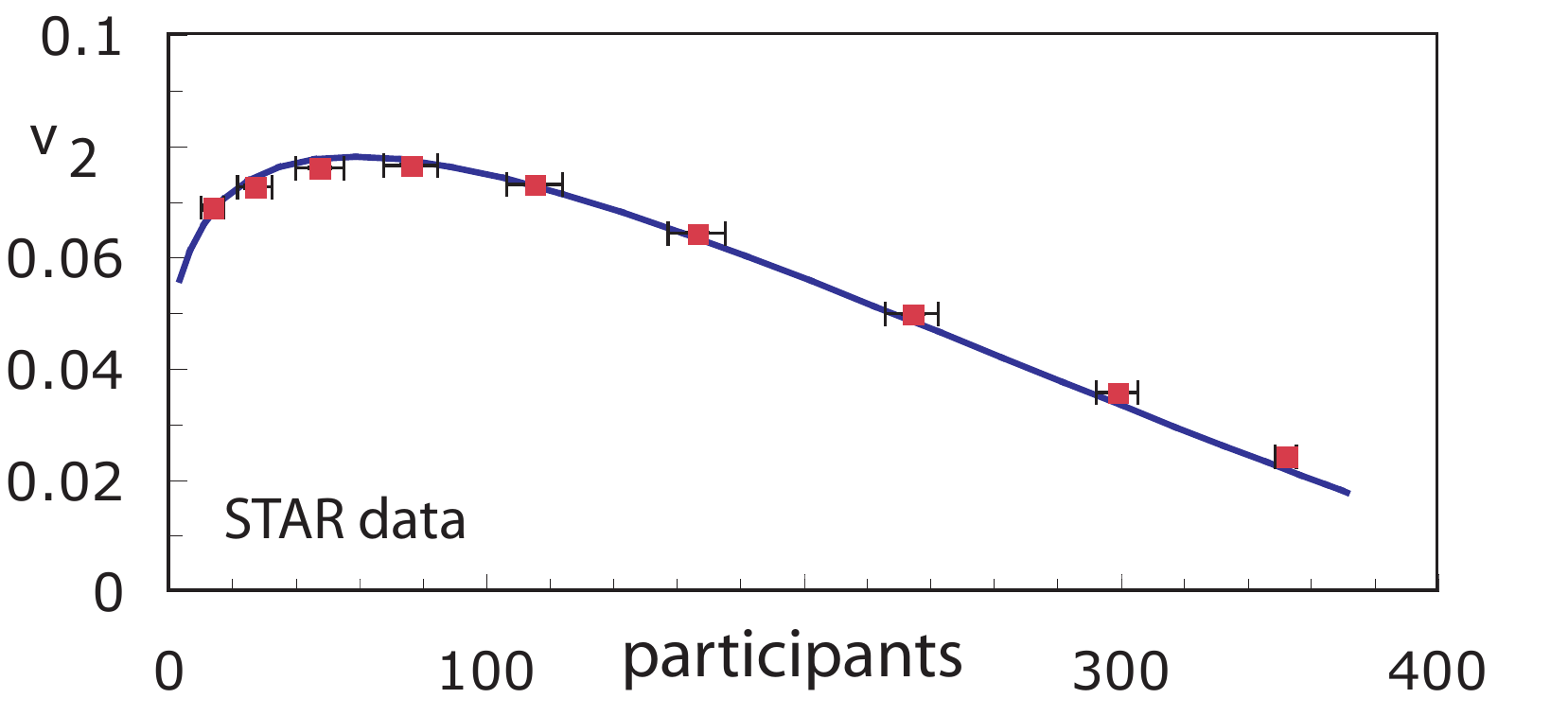}  \label{fig:1b}
\end{center}
\end{minipage}
\vspace{-1.5pc}
\caption{Average transverse momentum (left) and elliptic flow coefficient (right) compared to data from \cite{Adler:2003cb} and \cite{Adler:2002pu}.}
\end{figure}

The result shown in fig.~1 (right) agrees well with the data. Observe that viscous diffusion and transverse flow have the opposite effect on  $\sigma_\phi$. In the absence of flow, $\sigma_\phi$ would increase as a function of centrality due to diffusion. Consequently, information from $\sigma_\phi$ and $\sigma_\eta$ can constrain the viscosity-to-entropy ratio extracted from measurements of $\cal C$. Values in the range $0.08 < \eta/s < 0.11$ are consistent with the data in fig.~2, with other parameters held fixed. The problem remains, however, that $\Delta \sigma_{pt:n}^2$ is measured, not $\cal C$. The large uncertainty due to the possible role of density fluctuations remains. Direct measurements of $\cal C$ are needed to eliminate that uncertainty; until such data become available the estimate 
$0.08 < \eta/s < 0.32$ stands \cite{gavin06}.

{\em Note Added:} After the completion of this manuscript we learned of work  explaining the amplitude of the soft ridge as a consequence of  glasma correlations plus radial flow \cite{larry} . These authors do not explain the centrality dependence of  $\sigma_\phi$ and $\sigma_\eta$, while we do not address the amplitude.  We view these results aa complementary, since glasma may well determine the initial conditions for hydrodynamic evolution. 

This work was supported by a U.S.
National Science Foundation PECASE/CAREER award under grant
PHY-0348559.

\end{document}